# A VARIATIONAL DEDUCTION OF SECOND GRADIENT POROELASTICITY PART I: GENERAL THEORY

GIULIO SCIARRA, FRANCESCO DELL'ISOLA, NICOLETTA IANIRO AND ANGELA MADEO

Second gradient theories have to be used to capture how local micro heterogeneities macroscopically affect the behavior of a continuum. In this paper a configurational space for a solid matrix filled by an unknown amount of fluid is introduced. The Euler–Lagrange equations valid for second gradient poromechanics, generalizing those due to Biot, are deduced by means of a Lagrangian variational formulation. Starting from a generalized Clausius–Duhem inequality, valid in the framework of second gradient theories, the existence of a macroscopic solid skeleton Lagrangian deformation energy, depending on the solid strain and the Lagrangian fluid mass density as well as on their Lagrangian gradients, is proven.

## 1. Introduction

Poroelasticity stems from Biot's pioneering contributions on consolidating fluid saturated porous materials [Biot 1941] and now spans a lot of different interrelated topics, from geo- to biomechanics, wave propagation, transport, unsaturated media, etc. Many of these topics are related to modeling coupled phenomena (for example, chemomechanical swelling of shales [Dormieux et al. 2003; Coussy 2004], or biomechanical models of cartilaginous tissues), and nonstandard constitutive features (for instance, in freezing materials [Coussy 2005]). In all these cases, complexity generally remains in rendering how heterogeneities affect the macroscopic mechanical behavior of the overall material.

It is well known from the literature how microscopically heterogeneous materials can be described in the framework of statistically homogeneous media [Torquato 2002] considering suitable generalizations of the dilute approximation due to Eshelby [Nemat-Nasser and Hori 1993; Dormieux et al. 2006]; however, some lack in the general description of the homogenization procedure arises when dealing with heterogeneous materials, the characteristic length of which can be compared with the thickness of the region where high deformation gradients occur. This could be due, for example, to external periodic loading, the wavelength of which is comparable with the characteristic length of the material, or to phase transition, etc.

From the macroscopic point of view the quoted modeling difficulties, arising when high gradients occur, are discussed in the framework of so called high gradient theories [Germain 1973], where the assumption of locality in the characterization of the material response is relaxed. In these theories, the momentum balance equation reads in a more complex way than the classical one used for Cauchy continua. As a matter of fact, it is the divergence of the difference between the stress tensor and the divergence of so-called hyperstresses that balance the external bulk forces. Stress and hyperstress are introduced by a straightforward application of the principle of virtual power, as those quantities working on the gradient of velocity and the second gradient of velocity, respectively [Casal 1972; Casal and Gouin 1988]. Even the classical Cauchy theorem is, in this context, revised by introducing dependence of tractions not only on the outward normal unit vector but also on the local curvature of the boundary [dell'Isola and Seppecher 1997]; moreover symmetric and skew-symmetric couples (the actions called "double-forces" by Germain) must be prescribed on the boundary in terms of the hyperstress tensor together with contact edge forces along the lines where discontinuities of the normal vector occur.

Following the early papers on fluid capillarity [Casal 1972; Casal and Gouin 1988], the second gradient model can indeed be introduced by means of a variational formulation where the considered Helmholtz free energy depends both on the strain and the strain gradient tensors.

In the case of fluids, second gradient theories are typically applied for modeling phase transition phenomena [de Gennes 1985] or for modeling wetting phenomena [de Gennes 1985], when a characteristic length, say the thickness of a liquid film on a wall, becomes comparable with the thickness of the liquid/vapor interface [Seppecher 1993], annihilation (nucleation) of spherical droplets, when the radius of curvature is of the same order of the thickness of the interface [dell'Isola et al. 1996], or topological transition [Lowengrub and Truskinovsky 1998].

In the case of solids, second gradient theories are applied, for instance, when modeling the failure process associated with strain localization [Elhers 1992; Vardoulakis and Aifantis 1995; Chambon et al. 2004]. To the best of our knowledge, second gradient theories are very seldom applied in the mechanics of porous materials [dell'Isola et al. 2003] and no second gradient poromechanical model, consistent with the classical Biot theory, is available except the one presented in [Sciarra et al. 2007]. As gradient fluid models, second gradient poromechanics will be capable of providing significant corrections to the classical Biot model when considering porous media with characteristic length comparable to the thickness of the region where high fluid density (deformation) gradients occur. We refer, for instance, to crack/pore opening phenomena triggered by strain gradients or fluid percolation, the characteristic length being in this case the average length of the space between grains (pores).

Several authors have focused their attention on the development of homogenization procedures capable of rendering the heterogeneous response of the material at the microlevel by means of a second gradient macroscopic constitutive relation [Pideri and Seppecher 1997; Camar-Eddine and Seppecher 2003]; however, very few contributions seem to address this problem in the framework of averaging techniques [Drugan and Willis 1996; Gologanu and Leblond 1997; Koutzetzova et al. 2002]. The present work does not investigate the microscopic interpretation of second gradient poromechanics, but directly discusses its macroscopic formulation. It is divided into two papers: in the first paper the basics of kinematics, Section 2; the physical principles, Section 3; the thermodynamical restrictions, Section 4; and in Section 5 the variational deduction of the governing equations for a second gradient fluid filled porous material are presented.

In particular, in Section 2 a purely macroscopic Lagrangian description of motion is addressed by introducing two placement maps in $\chi_s$ and $\phi_f$ (Equation (1)). We do not explicitly distinguish which part of the current configuration of the fluid filled porous material is occupied at any time $t$ by the solid and fluid constituents, this information being partially included by the solid and fluid apparent density fields, which provide the density of solid/fluid mass with respect to the volume of the porous system (Equation (5)).

The deformation power, or stress working (Equation (12)), following Truesdell [1977] is deduced in Section 3 starting from the second gradient expression of power of external forces (Equation (9)) Cauchy theorem (Equation (10)) and balance of global momentum (see (11)).

In the spirit of Coussy et al. [1998] and Coussy [2004] thermodynamical restrictions on admissible constitutive relations are stated in Section 4, finding out a suitable overall potential, defined on the reference configuration of the solid skeleton. This last depends on the skeleton strain tensor and the fluid mass content, measured in the reference configuration of the solid, as well as on their Lagrangian gradients, in Equation (18).

Finally a deduction of the governing equations is presented in Section 5, based on the principle of virtual works, by requiring the variation of the internal energy to be equal to the virtual work of external and dissipative forces (see (19)). A second gradient extension of the two classical Biot equations of motion [Coussy 2004; Sciarra et al. 2007], endowed with the corresponding transversality conditions on the boundary, is therefore formulated (see Equations (30)–(33)). Generalizing the treatment developed, for example, by Baek and Srinivasa [2004] for first gradient theories, one of the equations of motion found by means of a variational principle is interpreted as the balance law for total momentum, when suitable definitions of the global stress and hyperstress tensors are introduced (see (34)).

In a subsequent paper (Part II, to be published in a forthcoming issue of this journal), an application of the second gradient model to the classical consolidation problem will be discussed. Our aim is to show how the present model enriches the description of a well-known phenomenon, typical of geomechanics, curing some of the weaknesses of the classical Terzaghi equation [von Terzaghi 1943]. In particular we will figure out the behavior of the fluid pressure during the consolidation process when varying the initial pressures of the solid skeleton and/or the saturating fluid. From the mathematical point of view, the initial boundary value problem will be discussed according with the theory of linear pencils.

## 2. Kinematics of fluid filled porous media and mass balances

The behavior of a fluid filled porous material is described, in the framework of a macroscopic model, adopting a Lagrangian description of motion with respect to the reference configuration of the solid skeleton. At any current time $t$ the configuration of the system is determined by the maps $\chi_s$ and $\phi_f$, defined as

$$\chi_s : \mathcal{B}_s \times \mathcal{I} \to \mathcal{E}, \qquad \phi_f : \mathcal{B}_s \times \mathcal{I} \to \mathcal{B}_f, \tag{1}$$

where $\mathcal{B}_\alpha$ ($\alpha = s, f$) is the reference configuration of the $\alpha$-th constituent, while $\mathcal{E}$ is the Euclidean place manifold, and $\mathcal{I}$ indicates a time interval. The map $\chi_s(\cdot, t)$ prescribes the current (time $t$) placement $\mathbf{x}$ of the skeleton material particle $\mathbf{X}_s$ in $\mathcal{B}_s$. The map $\phi_f(\cdot, t)$, on the other hand, identifies the fluid material particle $\mathbf{X}_f$ in $\mathcal{B}_f$ which, at time $t$, occupies the same current place $\mathbf{x}$ as the solid particle $\mathbf{X}_s$. Therefore the set of fluid material particles filling the solid skeleton is unknown, to be determined by means of evolution equations. Both these maps are assumed to be at least diffeomorphisms on $\mathcal{E}$. The current configuration $\mathcal{B}_t$ of the porous material is the image of $\mathcal{B}_s$ under $\chi_s(\cdot, t)$. In accordance with the properties of $\chi_s$ and $\phi_f$ it is straightforward to introduce the fluid placement map as

$$\chi_f : \mathcal{B}_f \times \mathcal{I} \to \mathcal{E}, \quad \text{such that } \chi_f(\cdot, t) = \chi_s(\cdot, t) \circ \phi_f(\cdot, t)^{-1},$$

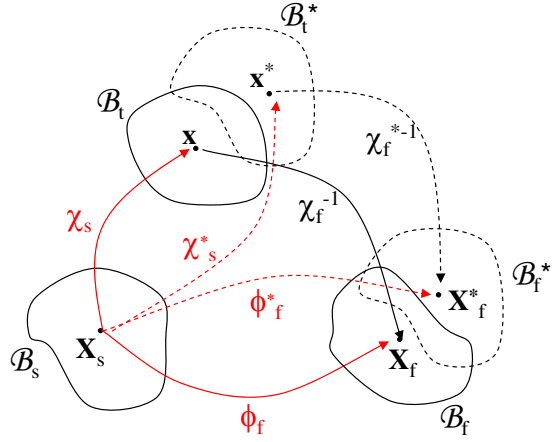

**Figure 1.** Lagrangian variations of the placement maps $\chi_s$, $\phi_f$, and $\chi_f$.

where $\chi_f(\cdot, t)$ is still a diffeomorphism on $\mathcal{E}$. Figure 1 shows how the introduced maps operate on the skeleton particle $X_s \in \mathcal{B}_s$; admissible variations of the two maps $\chi_s(\cdot, t)$ and $\phi_f(\cdot, t)$ are also depicted, in Section 5. In this way the space of configurations we will use has been introduced.

Independently of $t \in \mathcal{I}$, the Lagrangian gradients of $\chi_s$ and $\phi_f$ are introduced as

$$F_s(\cdot, t) : \mathcal{B}_s \to \operatorname{Lin}(V\mathcal{E}), \qquad \Phi_f(\cdot, t) : \mathcal{B}_s \to \operatorname{Lin}(V\mathcal{E}), \qquad (2)$$
$$X_s \mapsto \nabla_s \chi_s(X_s, t), \qquad\qquad X_s \mapsto \nabla_s \phi_f(X_s, t),$$

with $V\mathcal{E}$ being the space of translations associated to the Euclidean place manifold. In Equation (2) $\nabla_s$ indicates the Lagrangian gradient in the reference configuration of the solid skeleton; analogously, the gradient of $\chi_f$ is given by $F_f(X_f, t) = F_s(X_s, t) \cdot \Phi_f(X_s, t)^{-1}$, where $X_f = \phi_f(X_s, t)$. [1]

In the following the fluid Lagrangian gradient of $\chi_f$ will be indicated both by $F_f$ or $\nabla_f \chi_f$ when confusion can arise. Moreover, the time derivatives of $\chi_s$ and $\chi_f$, say the Lagrangian velocities of the solid skeleton and the fluid, can be introduced as

$$\text{for all } X_\alpha \in \mathcal{B}_\alpha, \quad \mathcal{V}_\alpha(X_\alpha, \cdot) : \mathcal{I} \to V\mathcal{E}, \quad t \mapsto \left. \frac{d\chi_\alpha}{dt} \right|_{(X_\alpha, t)}.$$

We also introduce the Eulerian velocities $v_\alpha$ as the push-forward of $\mathcal{V}_\alpha$ into the current domain

$$v_\alpha(\cdot, t) = \mathcal{V}_\alpha(\cdot, t) \circ \chi_\alpha(\cdot, t)^{-1}.$$

In the following we do not explicitly distinguish the map $\chi_s$ from its section $\chi_s(\cdot, t)$ if no ambiguity can arise. Moreover we will distinguish between the Lagrangian gradient ($\nabla_s$) in the reference configuration of the solid skeleton and the Eulerian gradient ($\nabla$) with respect to the current position $x$. Analogously, the solid Lagrangian and the Eulerian divergence operations will be noted by $\operatorname{div}_s$ and $\operatorname{div}$, respectively. All the classical transport formulas can be derived both for the solid and the fluid quantities; in particular,

---

[1] From now on we will indicate single, double and triple contraction between two tensors with $\cdot$, $:$, and $\vdots$ respectively.

those ones for an image volume and oriented surface element turn to be

$$d\mathcal{B}_t = J_\alpha d\mathcal{B}_\alpha, \qquad \boldsymbol{n} dS_t = J_\alpha \boldsymbol{F}_\alpha^{-T} . \boldsymbol{n}_\alpha dS_\alpha,$$

where $d\mathcal{B}_t$ and $dS_t$ represent the current elementary volume and elementary oriented surface corresponding to $d\mathcal{B}_\alpha$ and $dS_\alpha$, respectively, where $J_\alpha = \det \boldsymbol{F}_\alpha$, and where $\boldsymbol{n}$ and $\boldsymbol{n}_\alpha$ are the outward unit normal vectors to $dS_t$ and $dS_\alpha$. As far as only the solid constituent is concerned, we can understand that deformation induces changes in both the lengths of the material vectors and the angles between them. As it is well known, the Green–Lagrange strain tensor $\boldsymbol{\varepsilon}$ measures these changes, and is defined as

$$\boldsymbol{\varepsilon} := \tfrac{1}{2} \left( \boldsymbol{F}_s^T . \boldsymbol{F}_s - \boldsymbol{I} \right), \tag{3}$$

where $\boldsymbol{I}$ clearly represents the second order identity tensor.

The balance of mass both for the solid and the fluid constituent are introduced as

$$\mathbb{M}_\alpha = \int_{\mathcal{B}_t} \rho_\alpha \, d\mathcal{B}_t = \mathrm{const} = \int_{\mathcal{B}_\alpha} \rho_\alpha^0 \, d\mathcal{B}_\alpha, \quad (\alpha = s, f), \tag{4}$$

where $\mathbb{M}_\alpha$ is the total mass of the $\alpha$-th constituent, $\rho_\alpha$ is the current apparent density of mass of the $\alpha$-th constituent per unit volume of the porous material, while $\rho_\alpha^0$ is the corresponding density in the reference configuration of the $\alpha$-th constituent. When localizing, Equation (4) reads

$$\rho_\alpha J_\alpha = \rho_\alpha^0, \quad (\alpha = s, f),$$

or, in differential form,

$$\frac{d^\alpha \rho_\alpha}{dt} + \rho_\alpha \, \mathrm{div} \, (\boldsymbol{v}_\alpha) = 0, \quad (\alpha = s, f), \tag{5}$$

where $d^\alpha \rho_\alpha / dt$ represents the material time derivative relative to the motion of the $\alpha$-th constituent. In other words,

$$\frac{d^\alpha}{dt} := \left. \frac{d}{dt} \right|_{X_\alpha = \mathrm{const}}.$$

The macroscopic conservation laws could also be deduced in the framework of micromechanics [Dormieux and Ulm 2005; Dormieux et al. 2006] starting from a refined model, where the solid and the fluid material particles occupy two disjoint subsets of the current configuration, and considering an average of the solid and fluid microscopic mass balances. The macroscopic laws do involve the so called apparent density of the constituents and suitable macroscopic velocity fields. For a detailed description of the procedure which leads to averaged conservation laws we refer to the literature [Coussy 2004].

**2.1. *Pull back of continuity equations.*** It is clear that Equation (5) consists of Eulerian equations, meaning that they are defined on the current configuration of the porous medium. Following Wilmanski [1996] and Coussy [2004] we want to write both these equations in the reference configuration of the solid skeleton. With this purpose in mind let us define the relative fluid mass flow $\boldsymbol{w}$ as $\boldsymbol{w} := \rho_f (\boldsymbol{v}_f - \boldsymbol{v}_s)$. The use of this definition allows us to rearrange the fluid continuity (5) in the form

$$\frac{d^s \rho_f}{dt} + \rho_f \, \mathrm{div} \, \boldsymbol{v}_s + \mathrm{div} \, \boldsymbol{w} = 0. \tag{6}$$

We want now to rewrite the continuity equation for the fluid constituent in the reference configuration of the solid skeleton. The Lagrangian approach to the fluid mass balance can be carried out by introducing the current Lagrangian fluid mass content $m_f$, defined as

$$m_f := J_s \left( \rho_f \circ \chi_s \right). \tag{7}$$

Furthermore, let $\boldsymbol{M}$ be the Lagrangian vector referred to the reference configuration of the solid and related to the flow $\boldsymbol{w}$ through the relations

$$\boldsymbol{M} := J_s \boldsymbol{F}_s^{-1} . (\boldsymbol{w} \circ \chi_s), \qquad J_s \left( \text{div } \boldsymbol{w} \circ \chi_s \right) = \left( \text{div}_s \boldsymbol{M} \right). \tag{8}$$

By using the definitions from Equations (7) and (8) in (6) the fluid Lagrangian mass balance takes the form

$$\frac{dm_f}{dt} + \text{div}_s \boldsymbol{M} = 0.$$

## 3. Power of external forces

In this section, starting from the statement of the power of external forces for a second gradient solid-fluid mixture, we deduce its corresponding reduced form, accounting for the extended Cauchy theorem valid for second gradient continua [Casal 1972; Germain 1973; dell'Isola and Seppecher 1997], and the balance of global momentum. The external power $\mathcal{P}^{\text{ext}} \left( \boldsymbol{v}_s, \boldsymbol{v}_f \right)$ for a second gradient porous medium can be defined as a continuous linear functional of the velocity fields $\boldsymbol{v}_\alpha$; in particular

$$\mathcal{P}^{\text{ext}} \left( \boldsymbol{v}_s, \boldsymbol{v}_f \right) := \int_{\mathcal{B}_t} \left( \boldsymbol{b}_s . \boldsymbol{v}_s + \boldsymbol{b}_f . \boldsymbol{v}_f \right) d\mathcal{B}_t + \int_{\partial \mathcal{B}_t} \left( \boldsymbol{t}_s . \boldsymbol{v}_s + \boldsymbol{t}_f . \boldsymbol{v}_f \right) d\mathcal{S}_t + \int_{\partial \mathcal{B}_t} \left( \boldsymbol{\tau}_s . \frac{\partial \boldsymbol{v}_s}{\partial \boldsymbol{n}} + \boldsymbol{\tau}_f . \frac{\partial \boldsymbol{v}_f}{\partial \boldsymbol{n}} \right) d\mathcal{S}_t + \sum_{k=1}^{m} \int_{\mathcal{E}_k} \left( \boldsymbol{f}_s^k . \boldsymbol{v}_s + \boldsymbol{f}_f^k . \boldsymbol{v}_f \right) dl, \tag{9}$$

where $\mathcal{B}_t$ is the current volume occupied by the porous medium, $\partial \mathcal{B}_t$ its boundary, and $m$ is the number of edges $\mathcal{E}_k$ (if any) of the boundary. In addition, $\boldsymbol{b}_\alpha$, $\boldsymbol{t}_\alpha$, $\boldsymbol{\tau}_\alpha$, and $\boldsymbol{f}_\alpha^k$ represent the body force density, the generalized traction force (Cauchy stress vector), the double force vector, and the force per unit line acting on the $k$-th edge of the boundary, respectively.

The physical meaning of the double force $\boldsymbol{\tau}_\alpha$ can be described in a way similar to that used in different contexts in [Germain 1973] and [dell'Isola and Seppecher 1997]. It can be regarded as the sum of two different contributions, the first of which works on the rate of dilatancy along the outward unit normal $\boldsymbol{n}$ ($\nabla \boldsymbol{v}_\alpha : (\boldsymbol{n} \otimes \boldsymbol{n})$), and the second being a tangential couple working on the vorticity; this nomenclature is due to Germain [1973].

Let $\boldsymbol{\sigma}_\alpha$ and $\mathbb{C}_\alpha$ be the apparent Cauchy stress and hyperstress tensors per unit volume of the porous material relative to the $\alpha$-th constituent [Germain 1973; dell'Isola and Seppecher 1997]. The Cauchy theorem can be extended for a second gradient continuum, and in particular for a second gradient porous continuum, in order to specify how the generalized external tractions appearing in (9) can be balanced by the internal forces when considering any subdomain of the current volume. In particular we have[2]

$$t_\alpha = (\sigma_\alpha - \text{div}\,\mathbb{C}_\alpha).n - \text{div}^S(\mathbb{C}_\alpha.n), \qquad \tau_\alpha = (\mathbb{C}_\alpha.n).n, \qquad f_\alpha^k = [\![(\mathbb{C}_\alpha.n).\nu]\!]_k, \qquad (10)$$

where $\nu$ is the binormal unit vector which form a left-handed frame with the unit normal $n$ and the unit vector being tangent to the $k$-th edge. We note that, since $n$ is not continuous through the edge $k$, the vector $\nu$ is also discontinuous when passing from one side of the edge $k$ to the other. It is for this reason that the edge force $f_\alpha^k$ is balanced by the jump of the internal force $(\mathbb{C}_\alpha.n).\nu$ through the edge $k$. Extending classical poromechanics [Biot 1941; Coussy 2004] we now define the overall stress and hyperstress tensors as $\sigma := \sigma_f + \sigma_s$ and $\mathbb{C} := \mathbb{C}_f + \mathbb{C}_s$, so that the momentum balance for the porous medium as a whole reads

$$\text{div}\,(\sigma - \text{div}\,\mathbb{C}) + b = 0, \qquad (11)$$

where $b = b_s + b_f$ is the overall body force. Bearing in mind both the extended Cauchy theorem, Equation (10), and the overall balance of momentum, (11), together with the principle of virtual powers ($\mathcal{P}^{\text{ext}} = \mathcal{P}^{\text{def.}}$), (9) leads to the expression for the deformation power

$$\mathcal{P}^{\text{def.}}(v, \omega) = \int_{\mathcal{B}_t} \left[\sigma : \nabla v + \text{div}\left(\sigma_f^T.\omega\right) + \mathbb{C} \vdots \nabla\nabla v + \mathbb{C}_f \vdots \nabla\nabla\omega - \text{div}\,(\text{div}\,\mathbb{C}_f).\omega\right] d\mathcal{B}_t. \qquad (12)$$

Here and later on $v := v_s$ and $\omega := v_f - v_s$. Moreover, it must be remarked that (12) is obtained under the hypothesis of absence of volume forces ($b_\alpha = 0$) so that no inertia is taken into account in our model. We refer to [Coussy 2004] for the complete form of the deformation power in the case of first gradient porous continua. From now on, we also assume that the structure of the hyperstress tensors $\mathbb{C}_\alpha$ ($\alpha = f, s$) takes the particular form

$$\mathbb{C}_\alpha = I \otimes c_\alpha, \quad \alpha = f, s, \qquad (13)$$

where $I$ is the second order identity tensor and $c_\alpha$ is a kind of hyperstress vector related to the $\alpha$-th constituent. The use of this assumption restricts second gradient external forces just to vector fields $\tau_\alpha$, which only work on the stretching velocity of the $\alpha$-th constituent; in other words, no contribution to the vorticity on the boundary of $\mathcal{B}_t$ comes from $\tau_\alpha$. The aforementioned hypotheses indeed restrict the second gradient model; however, solid microdilatancies and capillarity effects can be still described by this second gradient model. According to (13) the external power due to second gradient effects, Equations (9) and (10), reduces to

$$\tau_\alpha.\frac{\partial v_\alpha}{\partial n} = \{[(I \otimes c_\alpha).n].n\}.\frac{\partial v_\alpha}{\partial n} = (c_\alpha.n)\,n.\frac{\partial v_\alpha}{\partial n} = (c_\alpha.n)\,[\nabla v_\alpha : (n \otimes n)].$$

**3.1. *Pull-back operations.*** Let us now consider the solid reference configuration pull-back of the deformation power; in order to do so, we will introduce the Piola–Kirchhoff like stress and hyperstress tensors for the overall body and for the fluid constituent. Thus, Piola–Kirchhoff stress ($S$) and hyperstress ($\gamma$) are defined so that

$$J_s \sigma : \nabla v =: S : \frac{d\varepsilon}{dt} \implies S = J_s F_s^{-1}.\sigma.F_s^{-T}, \qquad (14)$$

---

[2]Fixed a basis $(e_1, e_2, e_3)$, where $e_1$ and $e_2$ span the plane tangent to the surface $\partial\mathcal{B}_t$ at $x$, and the surface divergence of a second order tensor field $A$ is defined as $\text{div}^S A := \sum_{\alpha=1}^{2} (\partial A/\partial x_\alpha)\,e_\alpha$.

$$J_s \mathbb{C} \vdots \nabla\nabla \boldsymbol{v} =: \boldsymbol{\gamma} \cdot \left[ \left( \nabla_s \frac{d\boldsymbol{\varepsilon}}{dt} \right)^T : \boldsymbol{C}^{-1} - (\nabla_s \boldsymbol{C})^T : \left( \boldsymbol{C}^{-1} \cdot \frac{d\boldsymbol{\varepsilon}}{dt} \cdot \boldsymbol{C}^{-1} \right) \right] \implies \boldsymbol{\gamma} = J_s \boldsymbol{F}_s^{-1} \cdot \boldsymbol{c},$$

where $\boldsymbol{C} = \boldsymbol{F}_s^T \cdot \boldsymbol{F}_s$ is the Cauchy–Green strain tensor and $\boldsymbol{c}$ is the total hyperstress vector defined by $\boldsymbol{c} = \boldsymbol{c}_s + \boldsymbol{c}_f$. Moreover, the fluid ones are $\boldsymbol{S}_f =: J_s \boldsymbol{F}_s^{-1} \cdot \boldsymbol{\sigma}_f \cdot \boldsymbol{F}_s^{-T}$, and $\boldsymbol{\gamma}_f =: J_s \boldsymbol{F}_s^{-1} \cdot \boldsymbol{c}_f$. The deformation power $\mathcal{P}^{\text{def.}}$ can be finally written in the Lagrangian form

$$\mathcal{P}_{\mathcal{L}}^{\text{def.}} = \int_{\mathcal{B}_s} \hat{\mathcal{P}}_{\mathcal{L}}^{\text{def.}} d\mathcal{B}_s,$$

where

$$\hat{\mathcal{P}}_{\mathcal{L}}^{\text{def.}} = \left\{ \left[ \boldsymbol{S} - \boldsymbol{C}^{-1} \cdot ((\nabla_s \boldsymbol{C}) \cdot \boldsymbol{\gamma}) \cdot \boldsymbol{C}^{-1} \right] : \frac{d\boldsymbol{\varepsilon}}{dt} + (\boldsymbol{C}^{-1} \otimes \boldsymbol{\gamma}) \vdots \left( \nabla_s \frac{d\boldsymbol{\varepsilon}}{dt} \right) + \operatorname{div}_s \left( \frac{1}{m_f} \boldsymbol{S}_f^T \cdot \boldsymbol{M} \right) \right. $$
$$\left. + \nabla_s \left[ \operatorname{div}_s \left( \frac{\boldsymbol{M}}{m_f} \right) \right] \cdot \boldsymbol{\gamma}_f - \frac{\boldsymbol{M}}{m_f} \cdot \nabla_s \left( \operatorname{div}_s \boldsymbol{\gamma}_f \right) + \operatorname{div}_s \left[ \left( J_s^{-1} \frac{\boldsymbol{M}}{m_f} \cdot \nabla_s J_s \right) \boldsymbol{\gamma}_f \right] \right\}.$$

## 4. Thermodynamics: deduction of a macroscopic second gradient strain energy potential

In this section, starting from the first and second principles of thermodynamics, we will prove that a suitable macroscopic strain potential can be identified depending both on the solid strain and on the fluid mass density as well as on their Lagrangian gradients. Let $e_\alpha$ be the Eulerian density of internal energy relative to the $\alpha$-th constituent, and the corresponding energy density of the porous medium is defined as $e := \rho_s e_s + \rho_f e_f$. The first principle of thermodynamics can be written as [Coussy 2004]

$$\frac{d^s}{dt} \int_{\mathcal{B}_t} \rho_s e_s d\mathcal{B}_t + \frac{d^f}{dt} \int_{\mathcal{B}_t} \rho_f e_f d\mathcal{B}_t = \mathcal{P}^{\text{ext}} + \mathring{Q},$$

where $\mathring{Q} := -\int_{\mathcal{B}_t} \boldsymbol{q} \cdot \boldsymbol{n} \, d\mathcal{B}_t$ is the rate of heat externally supplied, and where $\boldsymbol{q}$ is the heat flow vector. In the Lagrangian form the first principle reads

$$\frac{d}{dt} \int_{\mathcal{B}_s} E \, d\mathcal{B}_s = \mathcal{P}_{\mathcal{L}}^{\text{def.}} - \int_{\mathcal{B}_s} \operatorname{div}_s \left( e_f \boldsymbol{M} + \boldsymbol{Q} \right) d\mathcal{B}_s. \tag{15}$$

where we recall that $d/dt$ is the material time derivative associated with the motion of the solid, $E := J_s e$ represents the Lagrangian density of internal energy, and $\boldsymbol{Q}$ is the Lagrangian heat flux defined by $\boldsymbol{Q} := J_s \boldsymbol{F}_s^{-1} \cdot \boldsymbol{q}$. Starting from Equation (15), the local Lagrangian form of the first principle is naturally given by

$$\frac{dE}{dt} = \hat{\mathcal{P}}_{\mathcal{L}}^{\text{def.}} - \operatorname{div}_s \left( e_f \boldsymbol{M} + \boldsymbol{Q} \right).$$

Let us now consider the second principle of thermodynamics and introduce the overall Eulerian density of entropy $s$ as $s := \rho_s s_s + \rho_f s_f$. The corresponding Lagrangian entropy is $S := J_s s$, and the Lagrangian form of the second principle can be written as [Coussy 2004]

$$\frac{d}{dt} \int_{\mathcal{B}_s} S \, d\mathcal{B}_s \geq - \int_{\mathcal{B}_s} \operatorname{div}_s \left( s_f \boldsymbol{M} + \frac{\boldsymbol{Q}}{T} \right) d\mathcal{B}_s. \tag{16}$$

If we now introduce the Helmholtz free energy $\Psi$ as $\Psi := E - TS$, Equation (16) can be rewritten in the local form as

$$\frac{dE}{dt} - S\frac{dT}{dt} - \frac{d\Psi}{dt} \geq -T \operatorname{div}_s \left( s_f \boldsymbol{M} + \frac{\boldsymbol{Q}}{T} \right).$$

Merging the local form of the first and the second principles, the extended Clausius–Duhem inequality (dissipation inequality) can be deduced. In particular, following Sciarra et al. [2007], we will distinguish different contributions to the dissipation function due to the solid and fluid motion and to thermal effects, respectively ($\Phi_s$, $\Phi_f$, and $\Phi_{th}$).

We now constitutively restrict [Coleman and Noll 1963] the admissible processes only to those ones which guarantee the dissipation inequality to be satisfied, because $\Phi_s$, $\Phi_f$ and $\Phi_{th}$ are separately nonnegative. In particular, the solid dissipation $\Phi_s$ reads

$$\Phi_s = \left\{ \boldsymbol{S} - \boldsymbol{C}^{-1}.[(\nabla_s \boldsymbol{C}).\boldsymbol{\gamma}].\boldsymbol{C}^{-1} - \left[ J_s \boldsymbol{C}^{-1}.\nabla_s \left( J_s^{-1} m_f \right) \right] \otimes \frac{\boldsymbol{\gamma}_f}{m_f} \right\} : \frac{d\boldsymbol{\varepsilon}}{dt} + \left( \boldsymbol{C}^{-1} \otimes \boldsymbol{\gamma} \right) \vdots \frac{d}{dt} \left( \nabla_s \boldsymbol{\varepsilon} \right)$$

$$+ \left[ g_f - \left( 1 + \frac{1}{tr \boldsymbol{I}} \right) \boldsymbol{\gamma}_f . \nabla_s \left( \frac{1}{m_f} \right) - \frac{J_s^{-1}}{tr \boldsymbol{I}} \frac{\boldsymbol{\gamma}_f}{m_f} . \nabla_s J_s \right] . \frac{dm_f}{dt} - \frac{\boldsymbol{\gamma}_f}{m_f} . \frac{d}{dt} \left( \nabla_s m_f \right) - S\frac{dT}{dt} - \frac{d\Psi}{dt}. \quad (17)$$

Assuming nondissipative processes occurring in the solid skeleton ($\Phi_s = 0$), Equation (17) allows for regarding the internal energy $\Psi$ as a state function

$$\Psi = \Psi \left( \boldsymbol{\varepsilon}, m_f, \nabla_s \boldsymbol{\varepsilon}, \nabla_s m_f, T \right). \quad (18)$$

From now on we will treat an isothermal problem and therefore assume the energy $\Psi$ does not depend on the temperature field $T$.

## 5. Variational deduction of second gradient poroelastic equations

**5.1.** *Basic concepts and first variation of the internal energy.* In this section we deduce the governing equations for a second gradient poroelastic continuum by means of a variational procedure. Variational approaches to first gradient mixture models are available in the literature [Bedford and Drumheller 1978; Gavrilyuk et al. 1998; Gouin and Ruggeri 2003].

In our case, we introduce the varied placement maps $\chi_s^*$ and $\phi_f^*$ for all $\boldsymbol{X}_s \in \mathcal{B}_s$ as

$$\chi_s^* (\boldsymbol{X}_s, t) = \chi_s (\boldsymbol{X}_s, t) + \delta \chi_s (\boldsymbol{X}_s, t), \qquad \phi_f^* (\boldsymbol{X}_s, t) = \phi_f (\boldsymbol{X}_s, t) + \delta \phi_f (\boldsymbol{X}_s, t),$$

where $\delta \chi_s$ and $\delta \phi_f$ represent arbitrary variations of the functions $\chi_s$ and $\phi_f$, respectively. The physical meaning of the variation $\delta \chi_s$ is well known in continuum mechanics, and stands for the virtual displacement (deformation) of the solid skeleton. The variation $\delta \phi_f$, instead, accounts for the virtual relative displacement of a fluid material particle with respect to a solid one (see Figure 1). Since these variations keep fixed $\boldsymbol{X}_s \in \mathcal{B}_s$ we label them Lagrangian variations and we note that the symbol $\delta$ commutes with the integral over $\mathcal{B}_s$ and with the Lagrangian gradient operator $\nabla_s$.

Following the statements of classical mechanics [Gantmacher 1970; Arnold 1989], the principle of virtual works reads

$$\delta \mathcal{A} = \delta \mathcal{L}^{\text{ext}} + \delta \mathcal{L}^{\text{diss}}, \quad (19)$$

where $\delta \mathcal{A}$ represents the Lagrangian variation of the internal energy of the porous material, defined as

$$\mathcal{A} := \int_{\mathcal{B}_s} \Psi \, d\mathcal{B}_s,$$

while $\delta \mathcal{L}^{\text{ext}}$ and $\delta \mathcal{L}^{\text{diss}}$ are the virtual works due to the external and dissipative forces acting on the porous system. Because of the aforementioned properties of the Lagrangian variations we can write

$$\delta \mathcal{A} = \delta \int_{\mathcal{B}_s} \Psi \, d\mathcal{B}_s = \int_{\mathcal{B}_s} \delta \Psi \, d\mathcal{B}_s. \tag{20}$$

Recalling now Equation (18), (20) implies

$$\delta \mathcal{A} = \int_{\mathcal{B}_s} \left( \frac{\partial \Psi}{\partial \boldsymbol{\varepsilon}} : \delta \boldsymbol{\varepsilon} + \frac{\partial \Psi}{\partial m_f} \delta m_f + \frac{\partial \Psi}{\partial (\nabla_s \boldsymbol{\varepsilon})} \vdots \delta (\nabla_s \boldsymbol{\varepsilon}) + \frac{\partial \Psi}{\partial (\nabla_s m_f)} . \delta (\nabla_s m_f) \right) d\mathcal{B}_s. \tag{21}$$

The variations $\delta \boldsymbol{\varepsilon}$, $\delta m_f$, $\delta (\nabla_s \boldsymbol{\varepsilon})$, and $\delta (\nabla_s m_f)$ must now be rewritten in terms of the variations of the primitive kinematical fields $\chi_s$ and $\phi_f$, bearing in mind that the Lagrangian variation commutes with the operator $\nabla_s$. We show here directly the results obtained in Appendix A, to which we refer for detailed calculations,

$$\delta \boldsymbol{\varepsilon} = \tfrac{1}{2} \left\{ (\nabla_s (\delta \chi_s))^T . \boldsymbol{F}_s + \boldsymbol{F}_s^T . \nabla_s (\delta \chi_s) \right\}, \tag{22}$$

and

$$\delta m_f = m_f \left( \nabla_s \phi_f \right)^{-T} : \nabla_s \left( \delta \phi_f \right). \tag{23}$$

Substituting (22) and (23) into (21), integration by parts (see Appendix B for details), allows us to write the variation of the second gradient potential $\mathcal{A}$ as

$$\delta \mathcal{A} = \int_{\mathcal{B}_s} A \, d\mathcal{B}_s + \int_{\partial \mathcal{B}_s} a \, d\mathcal{S}_s + \sum_{k=1}^{m} \int_{\mathfrak{E}_k} \alpha \, dl, \tag{24}$$

where $m$ is the number of edges $\mathfrak{E}_k$ of the body in the reference configuration of the solid and

$$A := - \text{div}_s \left[ \boldsymbol{F}_s . \left( \frac{\partial \Psi}{\partial \boldsymbol{\varepsilon}} - \text{div}_s \left( \frac{\partial \Psi}{\partial (\nabla_s \boldsymbol{\varepsilon})} \right) \right) \right] . \delta \chi_s + \left\{ (\nabla_s \phi_f)^{-T} . \left[ -m_f \nabla_s \left( \frac{\partial \Psi}{\partial m_f} - \text{div}_s \left( \frac{\partial \Psi}{\partial (\nabla_s m_f)} \right) \right) \right] \right\} . \delta \phi_f,$$

$$a := \left\{ \left[ \boldsymbol{F}_s . \left( \frac{\partial \Psi}{\partial \boldsymbol{\varepsilon}} - \text{div}_s \left( \frac{\partial \Psi}{\partial (\nabla_s \boldsymbol{\varepsilon})} \right) \right) \right] . \boldsymbol{n}_s - \text{div}_s^S \left[ \boldsymbol{F}_s . \left( \frac{\partial \Psi}{\partial (\nabla_s \boldsymbol{\varepsilon})} . \boldsymbol{n}_s \right) \right] \right\} . \delta \chi_s$$

$$+ \left\{ \left[ \boldsymbol{F}_s . \left( \frac{\partial \Psi}{\partial (\nabla_s \boldsymbol{\varepsilon})} . \boldsymbol{n}_s \right) \right] . \boldsymbol{n}_s \right\} . \frac{\partial (\delta \chi_s)}{\partial \boldsymbol{n}_s}$$

$$+ \left\{ (\nabla_s \phi_f)^{-T} . \left[ m_f \left( \frac{\partial \Psi}{\partial m_f} - \text{div}_s \left( \frac{\partial \Psi}{\partial (\nabla_s m_f)} \right) \right) \boldsymbol{n}_s - m_f \nabla_s^S \left( \frac{\partial \Psi}{\partial (\nabla_s m_f)} . \boldsymbol{n}_s \right) \right] \right\} . \delta \phi_f$$

$$+ \left[ (\nabla_s \phi_f)^{-T} . \left( m_f \frac{\partial \Psi}{\partial (\nabla_s m_f)} . \boldsymbol{n}_s \right) \boldsymbol{n}_s \right] . \frac{\partial (\delta \phi_f)}{\partial \boldsymbol{n}_s},$$

$$\alpha := \left\{ \left[ \boldsymbol{F}_s . \left( \frac{\partial \Psi}{\partial (\nabla_s \boldsymbol{\varepsilon})} . \boldsymbol{n}_s \right) \right] . \boldsymbol{\nu} \right\} . \delta \chi_s, + \left[ (\nabla_s \phi_f)^{-T} . \left( m_f \frac{\partial \Psi}{\partial (\nabla_s m_f)} . \boldsymbol{n}_s \right) \boldsymbol{\nu} \right] . \delta \phi_f.$$

**5.2. *Dissipative governing equations.*** In order to obtain the equations of motion for a second gradient poroelastic continuum, the form of the external and dissipation virtual works, $\delta\mathscr{L}^{\text{ext}}$ and $\delta\mathscr{L}^{\text{diss}}$, formally introduced in Equation (19), must be stated. The virtual dissipation $\delta\mathscr{L}^{\text{diss}}$ will account for the classical Darcy effects and for the so called Brinkman-like contributions [Brinkman 1947]. We define the dissipation $\delta\mathscr{L}^{\text{diss}}$ in the Eulerian configuration as

$$\delta\mathscr{L}^{\text{diss}} := -\int_{\mathscr{B}_t} \left\{ \mathbb{D}\left(\boldsymbol{v}_f - \boldsymbol{v}_s\right) . \left[\left(\delta\chi_f \circ \phi_f - \delta\chi_s\right) \circ \chi_s^{-1}\right] \right\} d\mathscr{B}_t$$

$$- \int_{\mathscr{B}_t} \left\{ \left[\mathbb{A} . \nabla\left(\boldsymbol{v}_f - \boldsymbol{v}_s\right)\right] : \nabla\left[\left(\delta\chi_f \circ \phi_f - \delta\chi_s\right) \circ \chi_s^{-1}\right] \right\} d\mathscr{B}_t, \quad (25)$$

where $\mathbb{D}$ is the symmetric, definite positive Darcy tensor and $\mathbb{A}$ is a suitably defined symmetric, definite positive second gradient Darcy-like tensor.

Moreover, from now on, we assume the following Eulerian expression for the external work $\delta\mathscr{L}^{\text{ext}}$,

$$\delta\mathscr{L}^{\text{ext}} := -\int_{\partial\mathscr{B}_t} \left\{ \boldsymbol{t} . \left(\delta\chi_s \circ \chi_s^{-1}\right) + \boldsymbol{t}_f . \left[\left(\delta\chi_f \circ \phi_f - \delta\chi_s\right) \circ \chi_s^{-1}\right] \right\} d\mathscr{S}_t. \quad (26)$$

We restrict our attention to $\boldsymbol{t}$ and $\boldsymbol{t}_f$, defined as

$$\boldsymbol{t} := p^{\text{ext}} \boldsymbol{n}, \qquad \boldsymbol{t}_f := \rho_f \mu^{\text{ext}} \boldsymbol{n}, \quad (27)$$

where $p^{\text{ext}}$ is the overall external pressure applied on $\partial\mathscr{B}_t$, and $\mu^{\text{ext}}$ is the chemical potential of the fluid outside the porous system. By comparison of Equation (26) with (9), we are assuming vanishing double forces and edge forces on the external boundary, as well as vanishing bulk actions. Equation (26), the expression for the external work, states that the external force $\boldsymbol{t}$ works only on the displacement of the solid skeleton ($\delta\chi_s$), while $\mu^{\text{ext}}$ works on the fluid mass virtual relative displacement $\rho_f\left(\delta\chi_f - \delta\chi_s\right)$. We note that if $\mu^{\text{ext}}$ is spatially constant then

$$\int_{\partial\mathscr{B}_t} \rho_f \mu^{\text{ext}} \left[\left(\delta\chi_f \circ \phi_f - \delta\chi_s\right) \circ \chi_s^{-1}\right] d\mathscr{S}_t = \int_{\partial\mathscr{B}_s} \left(\mu^{\text{ext}} \circ \chi_s\right) \delta m_f d\mathscr{S}_s,$$

that is, $\mu^{\text{ext}}$ works on the fluid mass which leaves (or enters) the solid skeleton (see Appendix C for details). Equation (26) for $\delta\mathscr{L}^{\text{ext}}$ can be rewritten (see Appendix C) on the reference configuration of the solid as

$$\delta\mathscr{L}^{\text{ext}} = \int_{\partial\mathscr{B}_s} \left\{ -\left(p^{\text{ext}} J_s \boldsymbol{F}_s^{-T} . \boldsymbol{n}_s\right) . \delta\chi_s + \left[\mu^{\text{ext}} m_f \left(\nabla_s \phi_f\right)^{-T} . \boldsymbol{n}_s\right] . \delta\phi_f \right\} d\mathscr{S}_s. \quad (28)$$

Finally, (25) for $\delta\mathscr{L}^{\text{diss}}$ (see Appendix C) assumes the Lagrangian form

$$\delta\mathscr{L}^{\text{diss}} = \int_{\mathscr{B}_s} \left\{ \left[\left(\nabla_s \phi_f\right)^{-T} . \left(J_s \boldsymbol{D} \boldsymbol{F}_s^T . \left(\mathscr{V}_f \circ \phi_f - \mathscr{V}_s\right)\right)\right] . \delta\phi_f \right\} d\mathscr{B}_s$$

$$- \int_{\mathscr{B}_s} \left\{ \left(\nabla_s \phi_f\right)^{-T} . \left[\boldsymbol{F}_s^T . \text{div}_s\left(J_s \mathbb{A} . \nabla_s \left(\mathscr{V}_f \circ \phi_f - \mathscr{V}_s\right) . (\boldsymbol{F}_s^T . \boldsymbol{F}_s)^{-1}\right)\right] \right\} . \delta\phi_f d\mathscr{B}_s$$

$$+ \int_{\partial\mathscr{B}_s} \left\{ \left[\left(\nabla_s \phi_f\right)^{-T} . \left(J_s \boldsymbol{F}_s^T . \mathbb{A} . \nabla_s \left(\mathscr{V}_f \circ \phi_f - \mathscr{V}_s\right) . (\boldsymbol{F}_s^T . \boldsymbol{F}_s)^{-1}\right)\right] . \boldsymbol{n}_s \right\} . \delta\phi_f d\mathscr{S}_s. \quad (29)$$

Starting from the principle of virtual works, (19), and using (24), (29), and (28) for $\delta \mathcal{A}$, $\delta \mathcal{L}^{\text{diss}}$, and $\delta \mathcal{L}^{\text{ext}}$, respectively, we can write the local equations of motion on $\mathcal{B}_s$ as

$$- \text{div}_s \left[ \boldsymbol{F}_s . \left( \frac{\partial \Psi}{\partial \boldsymbol{\varepsilon}} - \text{div}_s \left( \frac{\partial \Psi}{\partial (\nabla_s \boldsymbol{\varepsilon})} \right) \right) \right] = 0, \qquad (30)$$

and

$$(\nabla_s \phi_f)^{-T} . \left[ -m_f \nabla_s \left( \frac{\partial \Psi}{\partial m_f} - \text{div}_s \left( \frac{\partial \Psi}{\partial (\nabla_s m_f)} \right) \right) - J_s D \boldsymbol{F}_s^T . (\mathcal{V}_f \circ \phi_f - \mathcal{V}_s) \right]$$
$$+ (\nabla_s \phi_f)^{-T} . \left[ \boldsymbol{F}_s^T . \text{div}_s \left( J_s \mathbb{A} . \nabla_s (\mathcal{V}_f \circ \phi_f - \mathcal{V}_s) . (\boldsymbol{F}_s^T . \boldsymbol{F}_s)^{-1} \right) \right] = 0. \qquad (31)$$

Analogously the boundary conditions on $\partial \mathcal{B}_s$ read

$$\left[ \boldsymbol{F}_s . \left( \frac{\partial \Psi}{\partial \boldsymbol{\varepsilon}} - \text{div}_s \left( \frac{\partial \Psi}{\partial (\nabla_s \boldsymbol{\varepsilon})} \right) \right) \right] . \mathbf{n}_s - \text{div}_s^S \left[ \boldsymbol{F}_s . \left( \frac{\partial \Psi}{\partial (\nabla_s \boldsymbol{\varepsilon})} . \mathbf{n}_s \right) \right] = -J_s p^{ext} \boldsymbol{F}_s^{-T} . \mathbf{n}_s$$

$$(\nabla_s \phi_f)^{-T} . \left[ m_f \left( \frac{\partial \Psi}{\partial m_f} - \text{div}_s \left( \frac{\partial \Psi}{\partial (\nabla_s m_f)} \right) \right) \mathbf{n}_s - m_f \nabla_s^S \left( \frac{\partial \Psi}{\partial (\nabla_s m_f)} . \mathbf{n}_s \right) \right] +$$
$$- (\nabla_s \phi_f)^{-T} . \left\{ \left[ J_s \boldsymbol{F}_s^T . \mathbb{A} . \nabla_s (\mathcal{V}_f \circ \phi_f - \mathcal{V}_s) . (\boldsymbol{F}_s^T . \boldsymbol{F}_s)^{-1} \right] . \mathbf{n}_s \right\} = (\nabla_s \phi_f)^{-T} . (m_f \mu^{ext} \mathbf{n}_s), \quad (32)$$

$$\left[ \boldsymbol{F}_s . \left( \frac{\partial \Psi}{\partial (\nabla_s \boldsymbol{\varepsilon})} . \mathbf{n}_s \right) \right] . \mathbf{n}_s = 0,$$

$$(\nabla_s \phi_f)^{-T} . \left[ \left( m_f \frac{\partial \Psi}{\partial (\nabla_s m_f)} . \mathbf{n}_s \right) \mathbf{n}_s \right] = 0.$$

Finally, on the edges $\mathfrak{E}_k$ of the boundary (if any) the following conditions hold true:

$$\left[ \boldsymbol{F}_s . \left( \frac{\partial \Psi}{\partial (\nabla_s \boldsymbol{\varepsilon})} . \boldsymbol{n}_s \right) \right] . \boldsymbol{\nu} = \boldsymbol{0}, \qquad (\nabla_s \phi_f)^{-T} . \left[ \left( m_f \frac{\partial \Psi}{\partial (\nabla_s m_f)} . \boldsymbol{n}_s \right) \boldsymbol{\nu} \right] = \boldsymbol{0}. \qquad (33)$$

The Darcy and Brinkman dissipations appearing in Equations (31) and (32) can be rewritten in terms of the Lagrangian vector $\boldsymbol{M}$, previously defined as $\boldsymbol{M} = m_f \boldsymbol{F}_s^{-1} . (\boldsymbol{v}_f - \boldsymbol{v}_s)$. In fact, after some straightforward calculations, it can be proven that

$$\nabla (\boldsymbol{v}_f - \boldsymbol{v}_s) = \frac{1}{m_f} \left\{ [(\nabla_s \boldsymbol{F}_s)^T . \boldsymbol{M}]^T + \boldsymbol{F}_s . \nabla_s \boldsymbol{M} \right\} + \boldsymbol{F}_s . \left[ \boldsymbol{M} \otimes \nabla_s \left( \frac{1}{m_f} \right) \right].$$

We now show that (30) is in agreement with the classical second gradient balance law for the total momentum [Germain 1973; dell'Isola and Seppecher 1997]. In order to do so, considering assumption (13), it can be proven that the constitutive relations for $\boldsymbol{S}$ and $\boldsymbol{\gamma}$ (see Equations (14))

$$\frac{\partial \Psi}{\partial \boldsymbol{\varepsilon}} = \boldsymbol{S} - \boldsymbol{C}^{-1} . ((\nabla_s \boldsymbol{C}) . \boldsymbol{\gamma}) . \boldsymbol{C}^{-1}, \qquad \frac{\partial \Psi}{\partial (\nabla_s \boldsymbol{\varepsilon})} = \boldsymbol{C}^{-1} \otimes \boldsymbol{\gamma}, \qquad (34)$$

imply that (30) can be regarded as the solid-Lagrangian pull-back of (11). In other words,

$$J_s \, \text{div} \, (\boldsymbol{\sigma} - \text{div} \, \mathbb{C}) = \text{div}_s \left\{ \boldsymbol{F}_s . \left[ \boldsymbol{S} - \boldsymbol{C}^{-1} . ((\nabla_s \boldsymbol{C}).\boldsymbol{\gamma}).\boldsymbol{C}^{-1} - \boldsymbol{F}_s . \text{div}_s \left( \boldsymbol{C}^{-1} \otimes \boldsymbol{\gamma} \right) \right] \right\} = 0.$$

## 6. Concluding remarks

In this paper a purely macroscopic second gradient theory of poromechanics is presented, extending classical Biot poromechanics [Biot 1941; Coussy 2004]. Following a standard procedure, sketched in [Coussy et al. 1998], we determine a suitable representation formula of the deformation power, (12), for a second gradient porous medium, assuming the forces acting on solid skeleton to be balanced (using the generalized second gradient balance of momentum in the current domain and the generalized second gradient Cauchy theorem on its boundary) and the power of external forces to be that of two superposed second gradient continua [Germain 1973]. The principles of thermodynamics, together with the aforementioned representation of the deformation power, allow for deducing the existence of a suitable overall strain energy potential $\Psi$ depending on the solid strain tensor $\boldsymbol{\varepsilon}$ and the solid Lagrangian fluid mass density $m_f$, as well as on their Lagrangian gradients.

The Euler–Lagrange equations associated with the energy density $\Psi$ are the governing equations of the problem. In particular, Lagrangian variations of the placement maps $\chi_s$ and $\phi_f$ are considered. It is worth noting that the governing equations associated with the solid Lagrangian displacement $\delta \chi_s$ (when $\delta \phi_f = 0$) represents the balance of total momentum and therefore allows for the constitutive characterization of the overall stress and hyperstress tensors. This is a characteristic feature of the classical Biot model [Baek and Srinivasa 2004], which is completely recovered in this more general framework. On the other hand, the governing equation associated with the fluid placement map $\delta \phi_f$ represents the balance of momentum relative to the pure fluid, which, in this case, is a generalization of the classical Darcy law.

In part II, an application to the classical consolidation problem will show how the present model improve the classical ones. It is well known that second gradient theories are capable to detect boundary layer effects in the vicinity of interfaces; this is indeed what we will observe in the case of consolidation. In particular, a kind of fluid mass density increment in the neighborhood of the impermeable wall will be observed for the first time in a one dimensional problem [Mandel 1953; Cryer 1963].

## Appendix A: Basic variations

We show here how to derive the variations $\delta \boldsymbol{\varepsilon}$ and $\delta m_f$ in terms of the kinematical variations $\delta \chi_s$ and $\delta \phi_f$. Equation (3) for the Green–Lagrange strain tensor implies

$$\delta \boldsymbol{\varepsilon} = \tfrac{1}{2} \left[ \left( \delta_s \boldsymbol{F}_s^T \right).\boldsymbol{F}_s + \boldsymbol{F}_s^T.\delta_s \boldsymbol{F}_s \right],$$

where by definition $\boldsymbol{F}_s := \nabla_s \chi_s$; the expression (22) for $\delta \boldsymbol{\varepsilon}$ is easily derived. As far as the variation $\delta m_f$ is concerned, recalling definition (7) for $m_f$ we can write

$$m_f = J_s J_f^{-1} \rho_f^0, \tag{A.1}$$

where $\rho_f^0$ is the fluid density in the reference configuration of the fluid. Since by definition

$$J_f := \det(\nabla_f \chi_f) \quad \text{and} \quad \phi_f := \chi_f^{-1} \circ \chi_s,$$

we have
$$J_f = \det\left[\nabla_f\left(\chi_s \circ \phi_f^{-1}\right)\right] = \det\left(\nabla_s\chi_s . \nabla_f\phi_f^{-1}\right) = J_s \det\left(\nabla_s\phi_f\right)^{-1},$$

where for the sake of simplicity we neglect the dependence of the considered fields on the reference places. Equation (A.1) thus reads

$$m_f = \rho_f^0 \det\left(\nabla_s\phi_f\right). \tag{A.2}$$

By derivation rule of the determinant and assuming $\rho_f^0 =$ constant, we get

$$\delta m_f = \rho_f^0 \delta\left[\det\left(\nabla_s\phi_f\right)\right] = \rho_f^0 \det\left(\nabla_s\phi_f\right) \mathrm{tr}\left[\left(\nabla_s\phi_f\right)^{-1} . \delta\left(\nabla_s\phi_f\right)\right] = m_f\left[\left(\nabla_s\phi_f\right)^{-T} : \nabla_s\left(\delta\phi_f\right)\right].$$

## Appendix B: Variation of the internal energy

The procedure to calculate the variation $\delta\mathcal{A}$ of the internal energy will be here shown in detail.

According to Equations (21)–(23) and recalling that $\partial\Psi/\partial\boldsymbol{\varepsilon}$ is a symmetric second order tensor, while $\partial\Psi/\partial(\nabla_s\boldsymbol{\varepsilon})$ is a third order tensor symmetric with respect to its first two indices, we can write

$$\delta\mathcal{A} = \int_{\mathcal{B}_s}\left(A^1 + A_s^2 + A_f^2\right) d\mathcal{B}_s, \tag{B.1}$$

where

$$A^1 := \frac{\partial\Psi}{\partial\boldsymbol{\varepsilon}} : \left(\boldsymbol{F}_s^T.\nabla_s\left(\delta\chi_s\right)\right) + m_f\frac{\partial\Psi}{\partial m_f}\left(\nabla_s\phi_f\right)^{-T} : \nabla_s\left(\delta\phi_f\right),$$

$$A_s^2 := \frac{\partial\Psi}{\partial\left(\nabla_s\boldsymbol{\varepsilon}\right)} \vdots \nabla_s\left(\boldsymbol{F}_s^T.\nabla_s\left(\delta\chi_s\right)\right),$$

$$A_f^2 := \frac{\partial\Psi}{\partial\left(\nabla_s m_f\right)}.\nabla_s\left(m_f\left(\nabla_s\phi_f\right)^{-T} : \nabla_s\left(\delta\phi_f\right)\right),$$

account for the first gradient contribution to $\delta\mathcal{A}$ and for the solid and fluid second gradient contributions respectively.

The following identities are recalled in order to perform integrations by parts in (B.1); let $\lambda$, $\boldsymbol{a}$, $\boldsymbol{A}$, and $\mathbb{A}$ be scalar, first, second, and third order tensor fields respectively. (Here $\nabla^S\boldsymbol{a}$ indicates the surface gradient operator of a vector field $\boldsymbol{a}$ defined — analogously to $\mathrm{div}_s^S$ — as $\nabla^S\boldsymbol{a} = \partial\boldsymbol{a}/\partial x_\alpha \otimes \boldsymbol{e}_\alpha$, $\alpha = 1, 2$, where $\boldsymbol{e}_\alpha$ belong to the tangent plane.) Then,

$$\mathrm{div}\left(\boldsymbol{A}^T.\boldsymbol{a}\right) = \boldsymbol{A} : \nabla\boldsymbol{a} + \boldsymbol{a}.\,\mathrm{div}\,\boldsymbol{A}, \quad \mathrm{div}\left(\lambda\boldsymbol{A}\right) = \boldsymbol{A}.\nabla\lambda + \lambda\,\mathrm{div}\,\boldsymbol{A}, \quad \mathrm{div}\left(\lambda\boldsymbol{a}\right) = \boldsymbol{a}.\nabla\lambda + \lambda\,\mathrm{div}\,\boldsymbol{a},$$

$$\mathrm{div}\left(\mathbb{A}^T : \boldsymbol{A}\right) = \boldsymbol{A} : \mathrm{div}\,\mathbb{A} + (\nabla\boldsymbol{A}) \vdots \mathbb{A}, \quad \nabla\boldsymbol{a} = \nabla^S\boldsymbol{a} + \frac{\partial\boldsymbol{a}}{\partial\boldsymbol{n}}\otimes\boldsymbol{n},$$

where transposition for third order tensors is defined so as $\mathbb{A}^T := a_{ijk}\boldsymbol{e}_k\otimes\boldsymbol{e}_i\otimes\boldsymbol{e}_j$ if $\mathbb{A} = a_{ijk}\boldsymbol{e}_i\otimes\boldsymbol{e}_j\otimes\boldsymbol{e}_k$. Moreover, given second order tensors $\boldsymbol{A}$, $\boldsymbol{B}$, $\boldsymbol{C}$, third and first order tensors $\mathbb{A}$ and $\boldsymbol{a}$ the following identities are satisfied:

$$\boldsymbol{A} : \left(\boldsymbol{B}.\boldsymbol{C}\right) = \left(\boldsymbol{B}^T.\boldsymbol{A}\right) : \boldsymbol{C} = \left(\boldsymbol{A}.\boldsymbol{C}^T\right) : \boldsymbol{B}, \qquad \left(\mathbb{A}^T : \boldsymbol{A}\right).\boldsymbol{a} = \boldsymbol{A} : \left(\mathbb{A}.\boldsymbol{a}\right).$$

Finally, if $\varphi = \chi_s$ or $\varphi = \phi_f$, the identity holds true that

$$0 = \mathrm{div}_s \left[ \det (\nabla_s \varphi) (\nabla_s \varphi)^{-T} \right] = \det (\nabla_s \varphi) \, \mathrm{div}_s \left[ (\nabla_s \varphi)^{-T} \right] + (\nabla_s \varphi)^{-T} . \nabla_s \left[ \det (\nabla_s \varphi) \right]. \quad \text{(B.2)}$$

We underline that this equality holds unchanged for the surface divergence operator $\mathrm{div}_s^S$. For the sake of simplicity, we will perform integration by parts for the first and second gradient terms appearing in (B.1) separately. Integration by parts of the first gradient term, recalling Equation (A.2) for $m_f$ and using Equation (B.2) for $\phi_f$, leads to

$$\int_{\mathcal{B}_s} A^1 d\mathcal{B}_s = -\int_{\mathcal{B}_s} \mathrm{div}_s \left( F_s . \frac{\partial \Psi}{\partial \boldsymbol{\varepsilon}} \right) . \delta \chi_s \, d\mathcal{B}_s + \int_{\partial \mathcal{B}_s} \left[ \left( F_s . \frac{\partial \Psi}{\partial \boldsymbol{\varepsilon}} \right) . \boldsymbol{n}_s \right] . \delta \chi_s \, d\mathcal{S}_s$$
$$- \int_{\mathcal{B}_s} \left\{ (\nabla_s \phi_f)^{-T} . \left[ m_f \nabla_s \left( \frac{\partial \Psi}{\partial m_f} \right) \right] \right\} . \delta \phi_f \, d\mathcal{B}_s + \int_{\partial \mathcal{B}_s} \left[ (\nabla_s \phi_f)^{-T} . \left( m_f \frac{\partial \Psi}{\partial m_f} \boldsymbol{n}_s \right) \right] . \delta \phi_f \, d\mathcal{S}_s. \quad \text{(B.3)}$$

Integrating by parts the solid second gradient term we get

$$\int_{\mathcal{B}_s} A_s^2 d\mathcal{B}_s = -\int_{\mathcal{B}_s} \nabla_s (\delta \chi_s) : \left[ F_s . \mathrm{div}_s \left( \frac{\partial \Psi}{\partial (\nabla_s \boldsymbol{\varepsilon})} \right) \right] d\mathcal{B}_s + \int_{\partial \mathcal{B}_s} \nabla_s (\delta \chi_s) : \left[ F_s . \left( \frac{\partial \Psi}{\partial (\nabla_s \boldsymbol{\varepsilon})} . \boldsymbol{n}_s \right) \right] d\mathcal{S}_s$$
$$= -\int_{\partial \mathcal{B}_s} \left[ \left( F_s . \mathrm{div}_s \left( \frac{\partial \Psi}{\partial (\nabla_s \boldsymbol{\varepsilon})} \right) \right) . \boldsymbol{n}_s \right] . \delta \chi_s \, d\mathcal{S}_s + \int_{\mathcal{B}_s} \mathrm{div}_s \left[ F_s . \mathrm{div}_s \left( \frac{\partial \Psi}{\partial (\nabla_s \boldsymbol{\varepsilon})} \right) \right] . \delta \chi_s \, d\mathcal{B}_s$$
$$+ \int_{\partial \mathcal{B}_s} \left( \nabla_s^S (\delta \chi_s) + \frac{\partial (\delta \chi_s)}{\partial \boldsymbol{n}_s} \otimes \boldsymbol{n}_s \right) : \left[ F_s . \left( \frac{\partial \Psi}{\partial (\nabla_s \boldsymbol{\varepsilon})} . \boldsymbol{n}_s \right) \right] d\mathcal{S}_s.$$

Performing a further surface integration by parts we finally get

$$\int_{\mathcal{B}_s} A_s^2 d\mathcal{B}_s = -\int_{\partial \mathcal{B}_s} \left[ \left( F_s . \mathrm{div}_s \left( \frac{\partial \Psi}{\partial (\nabla_s \boldsymbol{\varepsilon})} \right) \right) . \boldsymbol{n}_s \right] . \delta \chi_s \, d\mathcal{S}_s + \int_{\mathcal{B}_s} \mathrm{div}_s \left[ F_s . \mathrm{div}_s \left( \frac{\partial \Psi}{\partial (\nabla_s \boldsymbol{\varepsilon})} \right) \right] . \delta \chi_s \, d\mathcal{B}_s$$
$$- \int_{\partial \mathcal{B}_s} \mathrm{div}_s^S \left( F_s . \left( \frac{\partial \Psi}{\partial (\nabla_s \boldsymbol{\varepsilon})} . \boldsymbol{n}_s \right) \right) . \delta \chi_s \, d\mathcal{S}_s + \int_{\partial \mathcal{B}_s} \left[ \left( F_s . \left( \frac{\partial \Psi}{\partial (\nabla_s \boldsymbol{\varepsilon})} . \boldsymbol{n}_s \right) \right) . \boldsymbol{n}_s \right] . \frac{\partial (\delta \chi_s)}{\partial \boldsymbol{n}_s} d\mathcal{S}_s$$
$$+ \sum_{k=1}^n \int_{\mathfrak{E}_k} \left[ \left( F_s . \left( \frac{\partial \Psi}{\partial (\nabla_s \boldsymbol{\varepsilon})} . \boldsymbol{n}_s \right) \right) . \boldsymbol{\nu} \right] . \delta \chi_s \, dl. \quad \text{(B.4)}$$

We finally rewrite the fluid second gradient term as

$$\int_{\mathcal{B}_s} A_f^2 d\mathcal{B}_s = \int_{\mathcal{B}_s} \frac{\partial \Psi}{\partial (\nabla_s m_f)} . \nabla_s \left[ m_f \, \mathrm{div}_s \left( (\nabla_s \phi_f)^{-1} . \delta \phi_f \right) \right] d\mathcal{B}_s$$
$$- \int_{\mathcal{B}_s} \frac{\partial \Psi}{\partial (\nabla_s m_f)} . \nabla_s \left[ m_f \, \mathrm{div}_s \left( (\nabla_s \phi_f)^{-T} \right) . \delta \phi_f \right] d\mathcal{B}_s;$$

recalling Equation (A.2) for $m_f$, using Equation (B.2) for $\phi_f$ and rearranging, we have

$$\int_{\mathcal{B}_s} A_f^2 \, d\mathcal{B}_s = \int_{\mathcal{B}_s} \frac{\partial \Psi}{\partial (\nabla_s m_f)} \cdot \nabla_s \left[ m_f \operatorname{div}_s \left( (\nabla_s \phi_f)^{-1} \cdot \delta \phi_f \right) \right] d\mathcal{B}_s$$
$$+ \int_{\mathcal{B}_s} \frac{\partial \Psi}{\partial (\nabla_s m_f)} \cdot \nabla_s \left[ \left( (\nabla_s \phi_f)^{-1} \cdot \delta \phi_f \right) \cdot \nabla_s m_f \right] d\mathcal{B}_s = \int_{\mathcal{B}_s} \frac{\partial \Psi}{\partial (\nabla_s m_f)} \cdot \nabla_s \left[ \operatorname{div}_s \left( m_f (\nabla_s \phi_f)^{-1} \cdot \delta \phi_f \right) \right] d\mathcal{B}_s.$$

Integrating by parts we get

$$\int_{\mathcal{B}_s} A_f^2 \, d\mathcal{B}_s = \int_{\partial \mathcal{B}_s} \operatorname{div}_s \left( m_f (\nabla_s \phi_f)^{-1} \delta \phi_f \right) \frac{\partial \Psi}{\partial (\nabla_s m_f)} \cdot \mathbf{n}_s \, d\mathcal{S}_s - \int_{\mathcal{B}_s} \operatorname{div}_s \left( m_f (\nabla_s \phi_f)^{-1} \delta \phi_f \right) \operatorname{div}_s \left( \frac{\partial \Psi}{\partial (\nabla_s m_f)} \right) d\mathcal{B}_s,$$

and, integrating again,

$$\int_{\mathcal{B}_s} A_f^2 \, d\mathcal{B}_s = \int_{\partial \mathcal{B}_s} m_f \operatorname{div}_s \left( (\nabla_s \phi_f)^{-1} \cdot \delta \phi_f \right) \left( \frac{\partial \Psi}{\partial (\nabla_s m_f)} \cdot \mathbf{n}_s \right) d\mathcal{S}_s$$
$$+ \int_{\partial \mathcal{B}_s} \left( \nabla_s m_f \cdot \left( (\nabla_s \phi_f)^{-1} \cdot \delta \phi_f \right) \right) \left( \frac{\partial \Psi}{\partial (\nabla_s m_f)} \cdot \mathbf{n}_s \right) d\mathcal{S}_s - \int_{\partial \mathcal{B}_s} \left[ m_f \operatorname{div}_s \left( \frac{\partial \Psi}{\partial (\nabla_s m_f)} \right) (\nabla_s \phi_f)^{-T} \cdot \mathbf{n}_s \right] \cdot \delta \phi_f \, d\mathcal{S}_s$$
$$+ \int_{\mathcal{B}_s} m_f \left[ (\nabla_s \phi_f)^{-T} \cdot \nabla_s \left( \operatorname{div}_s \left( \frac{\partial \Psi}{\partial (\nabla_s m_f)} \right) \right) \right] \cdot \delta \phi_f \, d\mathcal{B}_s.$$

Recalling again Equation (A.2) for $m_f$, using (B.2) for $\phi_f$ and rearranging we get

$$\int_{\mathcal{B}_s} A_f^2 \, d\mathcal{B}_s = \int_{\partial \mathcal{B}_s} \nabla_s (\delta \phi_f) : \left[ m_f \left( \frac{\partial \Psi}{\partial (\nabla_s m_f)} \cdot \mathbf{n}_s \right) (\nabla_s \phi_f)^{-T} \right] d\mathcal{S}_s$$
$$- \int_{\partial \mathcal{B}_s} \left[ m_f \operatorname{div}_s \left( \frac{\partial \Psi}{\partial (\nabla_s m_f)} \right) (\nabla_s \phi_f)^{-T} \cdot \mathbf{n}_s \right] \cdot \delta \phi_f \, d\mathcal{S}_s$$
$$+ \int_{\mathcal{B}_s} m_f \left[ (\nabla_s \phi_f)^{-T} \cdot \nabla_s \left( \operatorname{div}_s \left( \frac{\partial \Psi}{\partial (\nabla_s m_f)} \right) \right) \right] \cdot \delta \phi_f \, d\mathcal{B}_s.$$

Decomposing $\nabla_s\left(\delta\phi_f\right)$ as $\nabla_s\left(\delta\phi_f\right) = \nabla_s^S\left(\delta\phi_f\right) + \left(\partial\left(\delta\phi_f\right)/\partial\boldsymbol{n}_s\right)\otimes\boldsymbol{n}_s$, performing a last surface integration by parts and using (B.2) for the surface divergence operator we finally get

$$\int_{\mathcal{B}_s} A_f^2 d\mathcal{B}_s = \int_{\partial\mathcal{B}_s} \left\{\left(\nabla_s\phi_f\right)^{-T}\cdot\left[m_f\left(\frac{\partial\Psi}{\partial\left(\nabla_s m_f\right)}\cdot\boldsymbol{n}_s\right)\boldsymbol{n}_s\right]\right\}\cdot\frac{\partial\left(\delta\phi_f\right)}{\partial\boldsymbol{n}_s} d\mathcal{S}_s +$$
$$-\int_{\partial\mathcal{B}_s}\left\{\left(\nabla_s\phi_f\right)^{-T}\cdot\left[m_f\nabla_s^S\left(\frac{\partial\Psi}{\partial\left(\nabla_s m_f\right)}\cdot\boldsymbol{n}_s\right)\right]\right\}\cdot\delta\phi_f d\mathcal{S}_s + \sum_{k=1}^n\int_{\mathfrak{E}_k}\left\{\left(\nabla_s\phi_f\right)^{-T}\cdot\left[m_f\left(\frac{\partial\Psi}{\partial\left(\nabla_s m_f\right)}\cdot\boldsymbol{n}_s\right)\boldsymbol{\nu}\right]\right\}\cdot\delta\phi_f dl$$
$$-\int_{\partial\mathcal{B}_s}\left\{\left(\nabla_s\phi_f\right)^{-T}\cdot\left[m_f\operatorname{div}_s\left(\frac{\partial\Psi}{\partial\left(\nabla_s m_f\right)}\right)\boldsymbol{n}_s\right]\right\}\cdot\delta\phi_f d\mathcal{S}_s$$
$$+\int_{\mathcal{B}_s}\left\{\left(\nabla_s\phi_f\right)^{-T}\cdot\left[m_f\nabla_s\left(\operatorname{div}_s\left(\frac{\partial\Psi}{\partial\left(\nabla_s m_f\right)}\right)\right)\right]\right\}\cdot\delta\phi_f d\mathcal{B}_s. \quad \text{(B.5)}$$

Substituting (B.3), (B.4), and (B.5) into (B.1), the variation of the internal energy given in (24) has been recovered.

## Appendix C: External and dissipation works

The dissipation and external works have been defined in (25) and (26) on the Eulerian configuration of the system in terms of $\delta\chi_s$ and $\delta\chi_f\circ\phi_f$. These works must then be rewritten in terms of the independent variations $\delta\chi_s$ and $\delta\phi_f$. In order to do so, the relationship between $(\delta\chi_f\circ\phi_f - \delta\chi_s)$ and $\delta\phi_f$ must be established. We know by definition that $\chi_f\circ\phi_f = \chi_s$, so that $\delta\left(\chi_f\circ\phi_f\right) = \delta\chi_s$. Moreover, by differentiation rule for composite functions we have $\delta\chi_s = \delta\left(\chi_f\circ\phi_f\right) = \delta\chi_f\circ\phi_f + \left[\left(\nabla_f\chi_f\right)\circ\phi_f\right]\cdot\delta\phi_f$. But since $\chi_f = \chi_s\circ\phi_f^{-1}$, we get

$$\nabla_f\chi_f\circ\phi_f = \nabla_s\chi_s\cdot\left[\nabla_f\left(\phi_f^{-1}\right)\circ\phi_f\right] = \nabla_s\chi_s\cdot\left(\nabla_s\phi_f\right)^{-1},$$

so that $\delta\chi_s = \delta\chi_f\circ\phi_f + \nabla_s\chi_s\cdot\left(\nabla_s\phi_f\right)^{-1}\cdot\delta\phi_f$, or,

$$\delta\chi_f\circ\phi_f - \delta\chi_s = -\boldsymbol{F}_s\cdot\left(\nabla_s\phi_f\right)^{-1}\cdot\delta\phi_f. \quad \text{(C.1)}$$

We now prove that the external work due to the force $\boldsymbol{t}_f$ appearing in Equation (26) and prescribed by (27) represents the external work $\mathcal{L}_f^{\text{ext}}$ done to change the fluid mass inside the porous system when the external chemical potential $\mu^{\text{ext}}$ is assumed to be constant. We define this work as

$$\mathcal{L}_f^{\text{ext}} = \int_{\mathcal{B}_s}\left(\mu^{\text{ext}}\circ\chi_s\right)\delta m_f d\mathcal{B}_s;$$

according to (23) and neglecting composition operations we can write

$$\mathcal{L}_f^{\text{ext}} = \int_{\mathcal{B}_s}\mu^{\text{ext}} m_f\left(\nabla_s\phi_f\right)^{-T}:\nabla_s\left(\delta\phi_f\right) d\mathcal{B}_s,$$

which, integrating by parts, recalling (A.2) for $m_f$, and assuming $\mu^{\text{ext}}$ constant, gives

$$\mathscr{L}_f^{\text{ext}} = \int_{\partial\mathscr{B}_s} \left\{\left[\mu^{\text{ext}} m_f \left(\nabla_s \phi_f\right)^{-T} \cdot \boldsymbol{n}_s\right] \cdot \delta\phi_f\right\} d\mathscr{S}_s - \int_{\mathscr{B}_s} \mu^{\text{ext}} \rho_f^0 \operatorname{div}_s \left[\det\left(\nabla_s \phi_f\right) \left(\nabla_s \phi_f\right)^{-T}\right] d\mathscr{B}_s.$$

It is known from (B.2) that the divergence appearing in the second integral is vanishing, so that $\mathscr{L}_f^{\text{ext}}$ can be rewritten on the Eulerian configuration as

$$\mathscr{L}_f^{\text{ext}} = \int_{\partial\mathscr{B}_t} \left\{\left[\mu^{\text{ext}} \rho_f \left(\left(\nabla_s \phi_f\right)^{-T} \cdot \boldsymbol{F}_s^T\right) \cdot \boldsymbol{n}\right] \cdot \delta\phi_f\right\} \circ \chi_s^{-1} d\mathscr{S}_t,$$

or, using (C.1),

$$\mathscr{L}_f^{\text{ext}} = -\int_{\partial\mathscr{B}_t} \left[\rho_f \mu^{\text{ext}} \boldsymbol{n} \cdot \left(\delta\chi_f \circ \phi_f - \delta\chi_s\right)\right] \circ \chi_s^{-1} d\mathscr{S}_t,$$

which is the expression of the fluid external work used in (26).

The final expressions for $\delta\mathscr{L}^{\text{diss}}$ and $\delta\mathscr{L}^{\text{ext}}$ can now be determined. We first consider the solid Lagrangian pull-back of (25), which, recalling that $\nabla \boldsymbol{v}_\alpha = \nabla_s \mathscr{V}_\alpha \cdot \boldsymbol{F}_s^{-1}$, reads

$$\delta\mathscr{L}^{\text{diss}} := -\int_{\mathscr{B}_s} \left\{J_s D \left(\mathscr{V}_f \circ \phi_f - \mathscr{V}_s\right) \cdot \left[\left(\delta\chi_f \circ \phi_f - \delta\chi_s\right)\right]\right\} d\mathscr{B}_s$$
$$\qquad - \int_{\mathscr{B}_s} \left\{J_s \left[\mathbb{A} \cdot \nabla_s \left(\mathscr{V}_f \circ \phi_f - \mathscr{V}_s\right) \cdot \boldsymbol{F}_s^{-1}\right] : \nabla\left[\left(\delta\chi_f \circ \phi_f - \delta\chi_s\right)\right]\right\} d\mathscr{B}_s.$$

Recalling Equation (C.1), the dissipation work can be rewritten as

$$\delta\mathscr{L}^{\text{diss}} = \int_{\mathscr{B}s} \left[\left(\nabla_s \phi_f\right)^{-T} \cdot \left(J_s D \boldsymbol{F}_s^T \cdot \left(\mathscr{V}_f \circ \phi_f - \mathscr{V}_s\right)\right)\right] \cdot \delta\phi_f \, d\mathscr{B}_s$$
$$\qquad + \int_{\mathscr{B}s} \left\{J_s \left[\mathbb{A} \cdot \nabla_s \left(\mathscr{V}_f \circ \phi_f - \mathscr{V}_s\right) \cdot \boldsymbol{F}_s^{-1} \cdot \boldsymbol{F}_s^{-T}\right] : \nabla_s \left[\boldsymbol{F}_s \cdot \left(\nabla_s \phi_f\right)^{-1} \cdot \delta\phi_f\right]\right\} d\mathscr{B}_s;$$

integrating the second term by parts, Equation (29) for the dissipation work is easily recovered.

# References

[Arnold 1989] V. I. Arnold, *Mathematical methods of classical mechanics*, Springer Verlag, New York, 1989.

[Baek and Srinivasa 2004] S. Baek and A. R. Srinivasa, "Diffusion of a fluid through an elastic solid undergoing large deformation", *Int. J. Non-Linear Mech.* **39** (2004), 201–218.

[Bedford and Drumheller 1978] A. Bedford and D. S. Drumheller, "A variational theory of immiscible mixtures", *Arch. Rational Mech. Anal.* **68** (1978), 37–51.

[Biot 1941] M. A. Biot, "General theory of three-dimensional consolidation", *J. Appl. Phys.* **12** (1941), 155–164.

[Brinkman 1947] H. C. Brinkman, "A calculation of the viscous force exerted by a flowing fluid on a dense swarm of particles", *Appl. Sci. Res. A* **1** (1947), 27–34.

[Camar-Eddine and Seppecher 2003] M. Camar-Eddine and P. Seppecher, "Determination of the closure of the set of elasticity functionals", *Arch. Rational Mech. Anal.* **170** (2003), 211–245.

[Casal 1972] P. Casal, "La théorie du second gradient et la capillarité", *C. R. Acad. Sc. Paris Série A* **274** (1972), 1571–1574.

[Casal and Gouin 1988] P. Casal and H. Gouin, "Equations du mouvement des fluides thermocapillaires", *C. R. Acad. Sci. Paris Série II* **306** (1988), 99–104.

[Chambon et al. 2004] R. Chambon, D. Cailleire, and C. Tamagnini, "A strain space gradient plasticity theory for finite strain", *Comput. Methods Appl. Mech. Eng.* **193** (2004), 2797–2826.

[Coleman and Noll 1963] B. D. Coleman and W. Noll, "The thermodynamics of elastic material with heat conduction and viscosity", *Arch. Rational Mech. Anal.* **13** (1963), 167–178.

[Coussy 2004] O. Coussy, *Poromechanics*, J. Wiley & Sons, Chichester, 2004.

[Coussy 2005] O. Coussy, "Poromechanics of freezing materials", *J. Mech. Phys. Solids* **53** (2005), 1689–1718.

[Coussy et al. 1998] O. Coussy, L. Dormieux, and E. Detournay, "From mixture theory to Biot's approach for porous media", *Int. J. Solids Struct.* **35**:34–35 (1998), 4619–4635.

[Cryer 1963] C. W. Cryer, "A comparison of the three-dimensional consolidation theories of Biot and Terzaghi", *Q. J. Mech. Appl. Math.* **16**:4 (1963), 401–412.

[dell'Isola and Seppecher 1997] F. dell'Isola and P. Seppecher, "Edge contact forces and quasi balanced power", *Meccanica* **32** (1997), 33–52.

[dell'Isola et al. 1996] F. dell'Isola, H. Gouin, and G. Rotoli, "Nucleation of spherical shell-like interfaces by second gradient theory: numerical simulation", *Eur. J. Mech. B: Fluids* **15** (1996), 545–568.

[dell'Isola et al. 2003] F. dell'Isola, G. Sciarra, and R. C. Batra, "Static deformations of a linear elastic porous body filled with an inviscid fluid", *J. Elasticity* **72** (2003), 99–120.

[Dormieux and Ulm 2005] L. Dormieux and F. J. Ulm, *Applied micromechanics of porous materials series CISM n. 480*, Springer-Verlag, Wien New York, 2005.

[Dormieux et al. 2003] L. Dormieux, E. Lemarchand, and O. Coussy, "Macroscopic and micromechanical approaches to the modeling of the osmotic swelling in clays", *Transp. Porous Media* **50** (2003), 75–91.

[Dormieux et al. 2006] L. Dormieux, D. Kondo, and F. J. Ulm, *Microporomechanics*, Wiley, Chichester, 2006.

[Drugan and Willis 1996] W. J. Drugan and J. R. Willis, "A micromechanics-based nonlocal constitutive equation and estimates of representative volume element size for elastic component", *J. Mech. Phys. Solids* **44**:4 (1996), 497–524.

[Elhers 1992] W. Elhers, "An elastoplasticity model in porous media theories", *Transp. Porous Media* **9**:1–2 (1992), 49–59.

[Gantmacher 1970] F. R. Gantmacher, *Lectures in analytical mechanics*, MIR, Moscow, 1970.

[Gavrilyuk et al. 1998] S. Gavrilyuk, H. Gouin, and Y. V. Perepechko, "Hyperbolic models of homogeneous two-fluid mixtures", *Meccanica* **33** (1998), 161–175.

[de Gennes 1985] P. G. de Gennes, "Wetting: statics and dynamics", *Rev. Mod. Phys.* **57**:3 (1985), 827–863.

[Germain 1973] P. Germain, "La méthode des puissances virtuelles en mécanique des milieux continus", *J. Mecanique* **12**:2 (1973), 235–274.

[Gologanu and Leblond 1997] M. Gologanu and J. B. Leblond, *Recent extensions of Gurson's model for porous ductile materials in continuum micromechanics*, vol. 377, CISM Courses and Lectures, Springer, New York, 1997.

[Gouin and Ruggeri 2003] H. Gouin and T. Ruggeri, "Hamiltonian principle in the binary mixtures of Euler fluids with applications to second sound phenomena", *Rend. Mat. Acc. Lincei* **14**:9 (2003), 69–83.

[Koutzetzova et al. 2002] V. Koutzetzova, M. G. D. Geers, and W. A. M. Brekelmans, "Multi-scale constitutive modelling of heterogeneous materials with a gradient-enhanced computational homogenization scheme", *Int. J. Numer. Methods Eng.* **54** (2002), 1235–1260.

[Lowengrub and Truskinovsky 1998] J. Lowengrub and L. Truskinovsky, "Quasi-incompressible Cahn-Hilliard fluids and topological transitions", *Proc. R. Soc. Lond. A* **454** (1998), 2617–2654.

[Mandel 1953] J. Mandel, "Consolidation des sols (étude mathématique)", *Géothecnique* **3** (1953), 287–299.

[Nemat-Nasser and Hori 1993] S. Nemat-Nasser and M. Hori, *Micromechanics: Overall properties of heterogeneous materials*, North-Holland, Amsterdam, 1993.

[Pideri and Seppecher 1997] C. Pideri and P. Seppecher, "A second gradient material resulting from the homogenization of a heterogeneous linear elastic medium", *Continuum. Mech. Therm.* **9** (1997), 241–257.

[Sciarra et al. 2007] G. Sciarra, F. dell'Isola, and O. Coussy, "Second gradient poromechanics", *International Journal of Solids and Structures* **44** (2007), 6607–6629.

[Seppecher 1993] P. Seppecher, "Equilibrium of a Cahn-Hilliard fluid on a wall: influence of the wetting properties of the fluid upon the stability of a thin liquid film", *Eur. J. Mech. B:Fluids* **12** (1993), 169–184.

[von Terzaghi 1943] K. von Terzaghi, *Theoretical soil mechanics*, John Wiley & Sons, Chichester, 1943.

[Torquato 2002] S. Torquato, *Random heterogeneous materials*, Springer, New York, 2002.

[Truesdell 1977] C. A. Truesdell, *A first course in rational continuum mechanics*, Academic Press, New York, 1977.

[Vardoulakis and Aifantis 1995] I. Vardoulakis and E. C. Aifantis, "Heterogeneities and size effects in geo materials", *AMD - Am. Soc. Mech. Eng., Appl. Mech. Div. Newsl.* **201** (1995), 27–30.

[Wilmanski 1996] K. Wilmanski, "Porous media at finite strains. The new model with the balance equilibrium for porosity", *Arch. Mech.* **48**:4 (1996), 591–628.

GIULIO SCIARRA: giulio.sciarra@uniroma1.it
*Dipartimento di Ingegneria Chimica Materiali Ambiente, Università di Roma "La Sapienza", Via Eudossiana 18, 00184 Rome, Italy*

FRANCESCO DELL'ISOLA: francesco.dellisola@uniroma1.it
*Dipartimento di Ingegneria Strutturale e Geotecnica, Università di Roma "La Sapienza", Via Eudossiana 18, 00184 Rome, Italy*

and

*Laboratorio di Strutture e Materiali Intelligenti, Palazzo Caetani (Ala Nord), 04012 Cisterna di Latina (Lt), Italy*

NICOLETTA IANIRO: ianiro@dmmm.uniroma1.it
*Dipartimento di Metodi e Modelli Matematici per le Scienze Applicate, Università di Roma "La Sapienza", Via Scarpa 16, 00161 Rome, Italy*

ANGELA MADEO: angela.madeo@uniroma1.it
*Dipartimento di Metodi e Modelli Matematici per le Scienze Applicate, Università di Roma "La Sapienza", Via Scarpa 16, 00161 Rome, Italy*